\documentclass{iau_FM}
\usepackage{graphicx}

\title[S320.~~Solar Diameter from 1974 to 2015] 
{42 Years of Continuous Observations of the Solar Diameter from 1974 to 2015}

\author[A. H. Andrei, S. C. Boscardin, J. L. Penna, N. V. Leister \& C. Sigismondi]   
{A. H. Andrei, S. C. Boscardin, J. L. Penna,$^1$ \\ N. V. Leister,$^2$
 \and C. Sigismondi$^3$}

\affiliation{$^1$Observatorio Nacional, RJ, BR $^2$IAG/USP, SP, BR$^3$ICRA/Sapienza, Roma, IT 
email: {\tt sergio.boscardin@on.br , sigismondi@icra.it}}

\pubyear{2015}
\jname{Astronomy in Focus, Volume 1} 
\editors{Piero Benvenuti, ed.}
\begin{document}

\maketitle

\begin{abstract}
Several group in the World followed the solar diameter with dedicated instruments, namely solar astrolabes, since 1974. Their data have been gathered in several observing stations connected in the R2S3 (R\'eseau de Suivi au Sol du Rayon Solaire) network and through reciprocal visits and exchanges: Nice/Calern Observatory, Rio de Janeiro Observat\'orio Nacional/Brazil, IAG/Universidade de S\~ao Paulo/Brazil, Antalya Observatory/Turkey, San Fernando/Spain. The tradition of these observational efforts is here briefly sketched with the aim to evidence the possibility to analyze against the solar activity all these 42 years data at once by overcoming the problem of the shift between the different series. Each instrument has its own density filter with a prismatic effect responsible of that shift. The overall change of the solar radius during the last century is evident by comparing the Auwers' radius of 959.63'' (1891, present IAU standard) with 959.94'' (2015, from eclipses and Venus transit data, the latter either observed from space) and the role of ground-based daily measurements has been and it is crucial nowadays with Picard-sol and the Reflecting Heliometer of Rio de Janeiro, to ensure an homogeneus and continuous monitor of the solar diameter without the lifetime limitation of the satellites. 

\end{abstract}

\firstsection 
\section{Introduction: solar diameter measures from ground}

We present an analysis of 42 years of continuous measurements of the photospheric solar diameter, taken at major national observatories, using the same fundamental method, and similar apparatus. Such a series overlap observations from the Calern Observatory/France (Solar Astrolabe in 1975-2003 to 253 obs/year lead by F. Laclare and C. Delmas; Doraysol in 2000-2005 to 3,070 obs/year lead by C. Delmas and V. Sinceac), from the IAG/USP/Brazil (Solar Astrolabe in 1974-1994 to 95 obs/year lead by N. V. Leister, P. Benevides and M. Emilio), from the Antalya Observatory/Turkey (CCD Astrolabe in 2000-2007 to 400 obs/year lead by F. Chollet and O. Golbasi), from the San Fernando Observatory/Spain (Solar Astrolabe in 1972-1975 to 133 obs/year lead by J. Mui\~nos), from Observat\'orio Nacional/Brasil (CCD Astrolabe in 1998-2009 to 1,820 obs/year lead by J. Penna, E. Reis Neto and A.H. Andrei; Heliometer 2010-2015 to 8,509 obs/year lead by S.C. Boscardin, J.L. Penna and A.H. Andrei). The Heliometer is fully automatized in its observations and continues in regular operation with no plan of stopping; it shares with the former instruments the physical/mathematical definition of the limb, and the instruments aperture and focal length. We perform a reconciliation of all these series, using the common stretches. A modulation with the 11 years cycle of solar activity is evident. However when such modulation is removed, both from the solar diameter compound series and from the solar activity series (given by the sunspots count), a very strong anti-correlation surfaces. This suggests a smaller diameter for the forthcoming cycles, in a behavior similar to that on the Minima of Dalton (1810-14) and Maunder (1645-1715). This study stresses the importance of keeping and making available such long, continuous, and uniform series of solar diameter measurements. Maybe even the more by the controversy about its magnitude and origin. This presentation is dedicated to all the teams that developed and sustained the modern solar diameter observations. 

\section{Combined data from solar astrolabes 1974-2009}
In Nice/Calern observatory the Danjon Astrolabe operated from 1974 till 2007 (only some data after 2000) to partially overlap DORAYSOL observations(1999-2006, see figure 1 Calern CCD) with an average result of $959.48\pm0.01$, \cite[Morand, F. et al. (2010)]{Morand10}. 
In Valinhos (Brazil) the observations started in 1974, \cite[Leister and Benevides-Soares (1990)]{Leister and Benevides-Soares}.
The Reflecting Heliometer of Rio is operating since 2009 and it reached already 6 years of observations, 
\cite[Andrei et al. (2015)]{Andrei15} and \cite[Andrei et al. (2014)]{Andrei14}, with the intent to observe from ground the solar diameter in continuity with the past series of astrolabes, in particular in Rio the astrolabe operated from 1998 to 2009 (fig. 1),
with a partial overlap with the Heliometer sufficient to discover the prismatic effect of the solar density filters, \cite[Sigismondi, C.\etal\ (2015)].

\begin{figure}[t]
\begin{center}
 \includegraphics[width=5.2in]{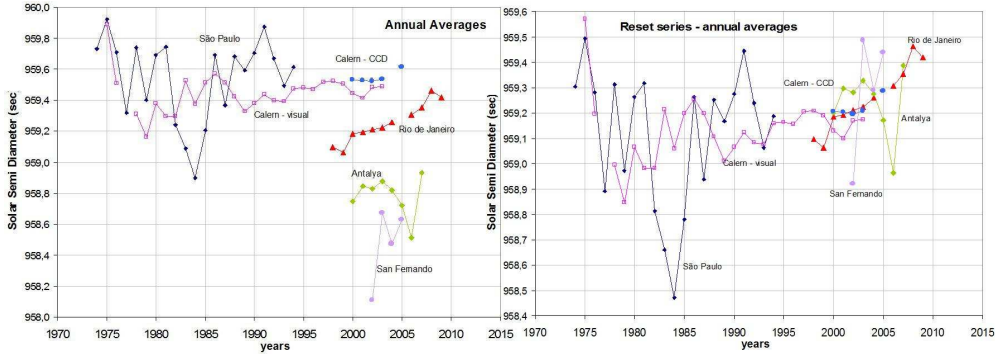} 
 \caption{All astrolabes data in a single plot. Different levels of each series are due to the filter's prismatic effect, \cite[Sigismondi et al. (2015)]{Sigismondi15}, with a general spread of 2'' in solar radius. Right side: reset averages with a final spread of all values reduced to 0.5''.}
   \label{fig1}
\end{center}
\end{figure}

\section{Solar diameter from space and ground: PICARD experiment}

The SODISM II instrument has been realized in Nice/Calern observatory to be calibrated with the spaceborn twin SODISM onboard PICARD satellite. 
The atmospheric parameters for SODISM II are measured with MISOLFA, \cite[Corbard et al. (2014)]{Corbard14}, 
a telescope with a front prism designed as the Heliometer of Schur and Ambronn in Goettingen Observatory during 1890-5, \cite[Meyermann (1939)]{Meyermann39}.    
The first results of PICARD satellite published on the solar diameter are already framed in a picture of a steady solar diameter, though limited to the rising phase of the cycle 24 \cite[Meftah, M. et al. (2015)]{Meftah15}. A series of papers started in 2000 on ApJ based on SOHO/MDI data were also entitled on the constancy of the solar diameter, see \cite[Bush, R. I., et al. (2010)]. Nevertheless the data of PICARD present space-based measured variations of the solar diameter within 20 mas in the 2010-2011 timespan, and ground-based within 50 mas in the 2011-2014 timespan.
The average of ground-based measures made by SODISM II is $959.86\pm0.18$ in 2011-2013 timespan \cite[Meftah, M. et al. (2014b)]{Meftah14b}.  
Finally the PICARD team claims also the constancy of the solar oblateness in the last two decades, measured from the
ground (e.g. Pic du Midi), balloons (SDS), and space (e.g. RHESSI). The equator-pole radius difference measured by PICARD is $8.4\pm0.5$ mas, which corresponds to an absolute radius difference of 6.1 km.

\section{Data from space: eclipse and transits}
An absolute and reference value was obtained by various teams with the transit of Venus in 2012. Using PICARD satellite data the radius at 607.1 nm resulted of $959.86\pm0.20$ arcsec \cite[Meftah, M. et al. (2014)]{Meftah14}, in agreement with the measures of the transit of Venus 2012 made with SDO/MDI by the same team [\cite[Hauchecorne et al. (2014)]]] of $959.90\pm0.06$ arcsec.
With the 10 cm full disk magnetograph 
at the Solar Observing Station of Huairou in China we also measured $959.91\pm0.16$ arcsec using the egress data of the 2012 Transit of Venus at 532.419 nm.
The measurements of the eclipses continued in the decade of 2010 with Adassuriya et al. for 2011 eclipse $959.89\pm0.19$ arcsec and with Lamy et al. 2015 who found as average of their values $959.99\pm0.06$ arcsec. 
Once again doing an average, from a statistical point of view, we are considering it as the most probable value, in the case of constancy of the parameter; 
for a variable Sun the average will change with the timespan upon which is calculated. The actual solar radius appears to be 0.3 arcsec larger than the previous IAU standard for the visible band: averaging the last eclipses data with the Venus transit the diameter is 959.9 arcsec vs 959.6 arcsec of Auwers (1891, with data including transits of Venus of 1874 and 1882). 
These data may suggest a larger diameter for the forthcoming cycles, a behavior possibly similar to that on the Minima of Dalton (1810-14) and Maunder (1645-1715). It is noteworthy to recall that several prior investigations show a direct correspondence between the short term variations of the solar diameter and the sunspots count (e.g., Boscardin et al., 2015, this meeting).

\section{Conclusions}
The current standard radius of 959.63'' was checked with several instruments and methods going back to 1874 transit of Venus, and in the last 42 years the same methods and instruments have been applied to new and older data obtaining an overall growth of the radius, even if nearly constant in the present decade of 2010, why to insist to consider constant something that is daily changing, and changed of 0.3 arcsec since the Auwers' radius of 959.63'' (1891) was chosen as a standard? 
The possibility to continue to monitor the solar radius is guaranteed by spaceborn instruments like SOHO, SDO, PICARD, RHESSI, but their lifetime is limited, and by ground-based instruments like the astrolabes, the Reflecting Heliometer, Picard-sol; the former have limited lifetime, the latter can be constantly upgraded and maintained.
The limit given by the clouds and the atmospheric turbulence is overcome by the potentially endless lifetime of the instruments, subjected only to rust and software ageing... things to be monitored by the astronomers and their administrations as routine work.

\end{document}